\documentclass[twocolumn,showpacs,preprintnumbers]{revtex4}

\usepackage{amssymb}
\usepackage{graphicx}
\usepackage{amsmath}
\usepackage{hyperref}

\begin{document}

\title{Theory of Spin Susceptibility in Frustrated Layered Antiferromagnets}

\author{A.F. Barabanov}
\email{abarab@bk.ru}

\affiliation{Institute for High Pressure Physics, Russian Academy
of Sciences, Troitsk 142190, Moscow Region, Russia}
\author{A.V. Mikheyenkov}
\affiliation{Institute for High Pressure Physics, Russian Academy
of Sciences, Troitsk 142190, Moscow Region, Russia}
\author{A.M. Belemuk}
\affiliation{Institute for High Pressure Physics, Russian Academy
of Sciences, Troitsk 142190, Moscow Region, Russia}

\date{\today}

\begin{abstract}
The self-consistent treatment of real and imaginary renormalizations in the
dynamic spin susceptibility $\chi (\mathbf{q},\omega )$ for the frustrated
Heisenberg model reproduces for cuprates at low doping: a spin spectrum $%
\omega _{\mathbf{q}}$, a saddle point for $\mathbf{q}\approx (\pi /2,\pi
/2), $ nearly constant $q$-integrated susceptibility $\chi _{2D}(\omega )$
for $\omega \lesssim 150\,meV$\ and a scaling law for $\chi _{2D}(\omega )$.
Frustration increase (optimally doped case) leads to a stripe scenario with
an $\omega _{\mathbf{q}}$-saddle point at $\mathbf{q\approx }(\pi ;\pi /2)$
and $\chi _{2D}(\omega )$\ peak at $\omega \approx 30\,meV$. The obtained $%
\chi (\mathbf{q},\omega )$ describes neutron scattering results and leads to
well-known temperature transport anomalies in doped cuprates.
\end{abstract}

\maketitle

The investigation of the dynamic spin susceptibility $\chi (\mathbf{q}%
,\omega )$ is a key problem for understanding the physics of layered high-$%
T_{c}$ superconductors (HTSC) in both low and optimally doped regimes.\ The
inelastic neutron scattering (INS) measurements in cuprates \cite{Lee2006}
-- \cite{Lake99} revealed a sharp resonant magnetic excitations peak of $%
\chi ^{^{\prime \prime }}(\mathbf{q},\omega )=\mathit{Im}\chi (\mathbf{q}%
,\omega )$ which corresponds to the aniferromagnetic (AFM) wave vector $%
\mathbf{Q}=(\pi ,\pi )$ at a resonant energy $E_{r}\approx 30\,meV$ and a
low-temperature peak at close energies for $q$-integrated susceptibility $%
\chi _{2D}(\omega ,T)$,\ $\chi _{2D}(\omega ,T)=\int d{\mathbf{q}}\chi
^{^{\prime \prime }}(\mathbf{q},\omega ,T)$. At low doping INS demonstrates
the scaling of magnetic response -- the universal law for $\chi _{2D}(\omega
,T)$ \cite{Keimer92} which states%
\begin{equation}
\frac{\chi _{2D}(\omega ,T)}{\chi _{2D}(\omega ,T\rightarrow 0)}=f(\omega /T)
\label{Scaling}
\end{equation}%
In the mentioned regime the spin excitation dispersion $\omega (\mathbf{q})$%
\ was measured across the entire Brillouin zone \cite{Coldea01} and it was
found that $\omega (\mathbf{q})$\ is anisotropic around the magnetic zone
boundary (a saddle point at $\mathbf{q\approx }(\pi ,\pi /2)$).

The aim of the work is to present a theory for the dynamic spin
susceptibility $\chi (\mathbf{q},\omega )$ within the frustrated $S=1/2$
Heisenberg model taking into account real and imaginary renormalizations
extracted from the irreducible Green's function $M(\mathbf{q,}\omega )$\ so
as to describe the mentioned experimental results in both doping regimes in
the framework of one self-consistent approach. Our analysis is based on a
spherically-symmetric treatment of the spin system which was introduced by
Shimahara and Takada \cite{st91} and generalized in \cite{Berezovsky}.

The recent microscopic theoretical progress in the investigations of $\chi (%
\mathbf{q},\omega )$ \cite{Prelovsek04, Larionov05, Sherman03} is based on $%
t-J$ model treated within the memory function approach (MFA). This approach
demonstrates $\chi ^{^{\prime \prime }}(\mathbf{Q},\omega )$\ peaks and the
scaling law. It is close to our treatment but it has difficulties in an
analytical calculation of the explicit form for$\ \omega (\mathbf{q})$\ and
a spin gap and as a result in the self-consistency procedure. Relative to
MFA our theory gives such new results as a demonstration of $\omega (\mathbf{%
q})$\ -- saddle point at $\mathbf{q\approx }(\pi ;\pi /2)$ and a new
analytical form for a scaling law for small frustration case (strongly
underdoped regime). For the large frustration (the regime close to optimal
doping) we reproduce not only $\chi ^{^{\prime \prime }}(\mathbf{Q,}\omega )$
peaks, but also the peaks of $\chi _{2D}(\omega )$ demonstrating a stripe
scenario (see \cite{Birgeneau06} for a review). For the latter case we
calculate also the resistivity $\rho (T)$ and the Hall coefficient $R(T)$ in
order to be sure that the found $\chi (\mathbf{q},\omega )$ reproduces the
well-known temperature anomalies in kinetics.

The Hamiltonian of the model has the form%
\begin{equation}
\widehat{H}_{I}=\frac{1}{2}I_{1}\sum_{\mathbf{i,g}}\overrightarrow{\mathbf{S}%
}_{\mathbf{i}}\overrightarrow{\mathbf{S}}_{\mathbf{i+g}}+\frac{1}{2}%
I_{2}\sum_{\mathbf{i,d}}\overrightarrow{\mathbf{S}}_{\mathbf{i}}%
\overrightarrow{\mathbf{S}}_{\mathbf{i+d}}  \label{Hamilt}
\end{equation}%
It describes the frustrated system of localized $S=1/2$ spins on a square
lattice, where $I_{1}$ is an AFM interaction constant for nearest, $I_{2}$
-- for next-nearest neighbors, $\mathbf{g,d}$ -- vectors of nearest and
next-nearest neighbors. We use standard variable $p$ (\textquotedblright
frustration parameter") $p=I_{2}/I$, $I_{1}=(1-p)I$, $I_{2}=pI,$ $%
I=I_{1}+I_{2}$ as a measure of frustration, hereinafter we treat all the
energetic parameters in the units of $I\ $and put $I=1$. We suppose that the
frustration (term $I_{2}$) simulates the influence of doping.

Following \cite{st91, Bar95,Berezovsky} we calculate $\chi (\mathbf{q}%
,\omega )=-\langle \langle S_{\mathbf{q}}^{\alpha }\mid ~S_{-\mathbf{q}%
}^{\alpha }\rangle \rangle _{\omega }$ -- two-time retarded Green's function
by the irreducible Green's function method. The dynamic spin susceptibility
can be written as $\ \chi (\mathbf{q},\omega )=-F_{\mathbf{q}}/(\omega
^{2}-\omega _{\mathbf{q}}^{2}-M(\mathbf{q},\omega )),$ where $\omega _{%
\mathbf{q}}$ is the spin excitations spectrum, $M(\mathbf{q},\omega
)=M^{^{\prime }}+iM^{\prime \prime }$ -- Fourier-transform of a new complex
three-site irreducible retarded Green's function, its analytical properties
are the same as those of $\chi (\mathbf{q},\omega )$.

Spectrum $\omega _{\mathbf{q}}$ and the numerator $F_{\mathbf{q}}$ have a
cumbersome form but they are expressed explicitly over five spin-spin
correlation functions $c_{\mathbf{g}},c_{\mathbf{d}},c_{\mathbf{g+g}},c_{%
\mathbf{d+d}},c_{\mathbf{g+d}}$ \cite{Berezovsky}, $c_{\mathbf{r}}=\langle
S_{\mathbf{R}}^{\alpha }S_{\mathbf{R+r}}^{\alpha }\rangle =(2\pi )^{-2}\int d%
\mathbf{q\,}c_{\mathbf{q}}e^{i\mathbf{qr}}$. This allows to write down and
solve numerically self-consistent system through the usual relations $c_{%
\mathbf{q}}=\left\langle S_{\mathbf{q}}^{z}S_{- \mathbf{q}}^{z}\right\rangle
=-\pi ^{-1}\int d\omega \,n_{B}(\omega )\mathit{Im}\langle \langle S_{%
\mathbf{q}}^{\alpha }\mid S_{-\mathbf{q}}^{\alpha }\rangle \rangle _{\omega
+i\delta }$, $\ S_{\mathbf{q}}^{\alpha }=N^{-1/2}{\,}\sum\limits_{\mathbf{q}%
}e^{-i\mathbf{qr}}S_{\mathbf{r}}^{\alpha }$. The set of equations includes
also the sum-rule condition $c_{\mathbf{r=0}}=1/4$. The system is solved at
every fixed $T$ and $p$.

The imaginary part $M^{\prime \prime }(\mathbf{q},\omega )$ is an odd
function of $\omega $. In the simplest approach \cite{Prelovsek04}\ we put $%
M^{\prime \prime }(\mathbf{q},\omega )=-\omega \gamma $, where the damping $%
\gamma $ is taken to be independent on $\mathbf{q}$ and $\omega $. We take
the real part $M^{^{\prime }}$ as $M^{^{\prime }}\sim \left\vert \sin
(q_{x})\sin (q_{y})\right\vert ^{3}$ and introduce a renormalized spectrum $%
\widetilde{\omega }_{\mathbf{q}}^{2}=\omega _{\mathbf{q}}^{2}+(\lambda |\sin
q_{x}\sin q_{y}|)^{3}$. The choice of $M^{^{\prime }}$functional form is
dictated by the condition that $M^{^{\prime }}$ represents the square
harmonic different from those involved in the functional form of $\omega
^{2}(\mathbf{q})$. Though the $\lambda $-renormalization is zero along the
lines $\mathbf{\Gamma -X}$ and $\mathbf{X-M}$ ($\mathbf{\Gamma =}(0,0),$ $%
\mathbf{X=}(0,\pi ),$ $\mathbf{M=}(\pi ,\pi )$) and mainly modifies the top
of the spectrum, it changes the spin gap $\Delta _{\mathbf{M}}=\widetilde{%
\omega }_{\mathbf{Q}}$\ due to self-consistency of calculations. So the
dynamic spin susceptibility%
\begin{equation}
\chi (\mathbf{q},\omega )=\frac{-F_{\mathbf{q}}}{\omega ^{2}-\widetilde{%
\omega }_{\mathbf{q}}^{2}+i\omega \gamma },  \label{hi_our}
\end{equation}%
contains two parameters $\gamma $ and $\lambda $.

We relate the dielectric limit to the case of extremely small frustration $%
p=0.04$. In the inset of Fig.1 the spectrum $\widetilde{\omega }(\mathbf{q})$
is presented in this limit for $T=0.1$, $\gamma =0.025$ and $\lambda =-1.0$ (%
$T\sim 100K$ for $I\sim 100\,meV$). The spectrum is almost linear on $%
\widetilde{q}=\left\vert \mathbf{q}-\mathbf{Q}\right\vert $ up to $\omega
_{0}\sim 1.5$. It can be found that for fixed $\mathbf{q}$ there is a
well-defined $\chi (\mathbf{q},\omega )$ peak on $\omega $\ which is
related\ to the spectrum $\widetilde{\omega }(\mathbf{q})$. More exactly,
the maximum of $\chi (\mathbf{q},\omega )$ on $\omega $\ corresponds to the
frequency close to $\widetilde{\omega }(\mathbf{q}),$\ but always a bit
smaller (due to damping $\gamma )$. For $I=1.2\ meV$\ a spin-wave velocity $%
\hbar c\approx 900$ $meV{\,}\mathring{A}$\ is close to the value given in
\cite{Hayden96-PRL}. As it is seen from the inset of Fig.1, in accordance
with the experiments \cite{Coldea01},\ the dispersion $\widetilde{\omega }(%
\mathbf{q})$ is anisotropic around the magnetic zone boundary and has a
saddle point close to $\mathbf{q=Q}/2$ ($\widetilde{\omega }(\mathbf{q=}%
(0,\pi )>\widetilde{\omega }(\mathbf{q=}(\pi /2,\pi /2))$). Note that in
contrast to our treatment one needs to adopt a ferromagnetic second-neighbor
exchange $I_{2}$ with $p\leq -0.1$\ for the explanation of such an
anisotropy in the framework of the linear spin-wave theory \cite{Coldea01}.

\begin{figure}
\includegraphics[width=8cm]{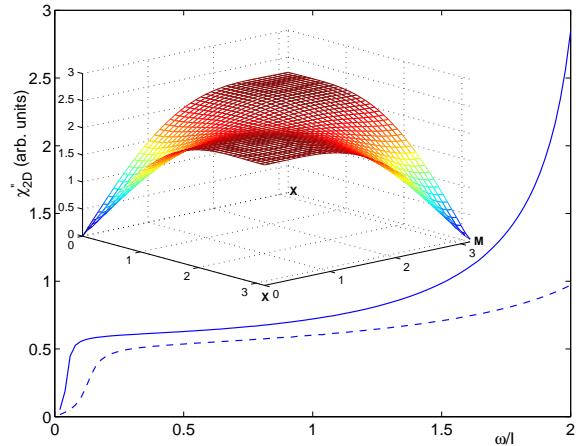}
\caption
{$\chi _{2D}(\omega )$ for frustration $p=0.04$: $T/I=0.1$ -- solid, $T/I=0.3 $
-- dashed curve (damping $\gamma =0.5T$). {\it Inset}:
self-consistent spectrum $\widetilde{\omega }(\mathbf{q})$ for
$p=0.04$ and $T/I=0.1$}
\label{fig:fig1}
\end{figure}

In Fig.1 $\chi _{2D}(\omega )$ is given for $p=0.04$, $\gamma =0.25T$\ in
two cases: $T=0.1$, $\lambda =-1.0$\ and $T=0.3$, $\lambda =2.0^{1/3}$. The $%
\lambda $-values are chosen from the condition that the resulting spin gap
should be approximately linear on $T$\ : $\Delta _{\mathbf{M}}(T=0.1)=0.048$%
, $\Delta _{\mathbf{M}}(T=0.3)=0.134$. Below we show analytically that in
the low-frustration limit this linearity is the necessary condition for the
scaling law. The qualitative coincidence of calculated function with the
experiment \cite{Hayden96-PRL,Hayden96-PR} is seen -- $\chi _{2D}(\omega )$
is nearly constant in a large $\omega $ interval and increases for $\omega
>150\,meV$.

\begin{figure}
\includegraphics[width=8cm]{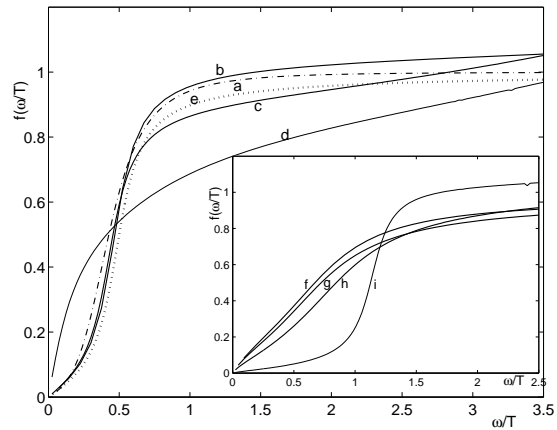}
\caption {Scaling curves $f(\omega /T)$ for $p=0.04$: the
dashed-dotted line \textbf{a} -- best fit for scaling in
$La_{1.96}Sr_{0.04}CuO_{4}$ \cite{Keimer92}; solid lines
\textbf{b} and \textbf{c} were calculated at $T/I=0.1;$ $0.3$
for $\gamma =0.25T$; dotted line \textbf{e} -- scaling law (\ref{fi_from_hi})
-- see text; solid thin curve {\textbf{d}}-- destroyed scaling $T=0.3$ and $\gamma /I=0.3>0.25T/I=0.075$.
{\it Inset}: Scaling for $p=0.1$: solid curves {\textbf{f},\textbf{g},\textbf{h}}
respectively for $T/I=0.1;$$0.2;$$0.4$, $\gamma =T$. The solid thin curve \textbf{l }-- $T/I=0.4$ and
$\gamma /I=0.1<T/I=0.4.$}
\label{fig:fig2}
\end{figure}

Now we treat the scaling condition which leads to a strong limitation of $%
\gamma (T)$ dependence. Fig.2 represents the scaling functions $f(\omega /T)$%
. Solid lines \textbf{b} and \textbf{c} correspond to temperatures $T=0.1$
and $T=0.3$\ respectively and are calculated for the same parameters, as in
Fig.1 (that is, in particular, for $\gamma =0.25T$). The dash-dotted line
\textbf{a} is the best fit for experimental scaling in $%
La_{1.96}Sr_{0.04}CuO_{4}$ \cite{Keimer92} $f_{ex}(\omega /T)=(2/\pi
)\arctan \{0.43(\omega /T)+10.5(\omega /T)^{3}\}$. It is approximately a
step function on $(\omega /T)$\ smeared through $\delta =\Delta (\omega
/T)\simeq 0.25$. The calculated curves have close value of $\delta $. Note
that the value of $\delta $ strongly restricts the $\gamma (T)$ dependence.
For example,\ thin curve \textbf{d}\ in Fig.2 corresponds to $T=0.3$, $%
\lambda =2.0^{1/3}$ and $\gamma =0.3>0.25T=0.075$. As a result curve \textbf{%
d} strongly deviates from curves \textbf{a}\ ($\ f_{ex}$ ) and \textbf{c}\ ($%
f_{T=0.3,\gamma =0.25T})$ and it has $\delta \gg 0.25$.

The analogous picture for the frustration $p=0.1$ is shown in the inset of
Fig.2 for $\gamma =T$ and $\lambda =0$ We relate this case to strongly
underdoped $Y$\textbf{-}cuprates \cite{Stock04}. The calculated scaling
functions are given for the temperatures $T=0.1,0.2,0.4$ by the solid curves
\textbf{f}, \textbf{g}, \textbf{h} respectively. It can be seen that they
are close to the experimental $f_{ex}(\omega /T)=(\frac{2}{\pi })\arctan
\{a(\omega /T)\}$, $a\sim 1$ \cite{Stock04}. The thin curve \textbf{l}
corresponds to $T=0.4$ and $\gamma =0.1<T=0.4$. Comparison of the curves
\textbf{h} and \textbf{l} explicitly demonstrates that, as in the previous
case $p=0.04$ ,\ the scaling is destroyed when $\gamma (T)$ deviates from a
linear law.

Thus, the above results demonstrate that the scaling law holds if $\gamma $
is a linear function on $T$. In the limit $p\ll 1$ this point can be
clarified analytically taking $\mathit{Im}\chi (\mathbf{q},\omega )=$ $%
\gamma \omega F_{\mathbf{q}}/\{(\omega ^{2}-\omega _{\mathbf{q}%
}^{2})^{2}+\gamma ^{2}\omega ^{2}\}$ It is obvious from the inset of Fig.1
that for $\omega /T\leq 2,~T/I\leq 0.3$\ the main input to $\chi
_{2D}(\omega )$ is given by the region $\widetilde{q}\leq \widetilde{q}_{0}$%
; $c\widetilde{q}_{0}\sim I$ is the largest energetic parameter. Then $%
\omega ^{2}(q)\approx \Delta _{\mathbf{M}}^{2}+c^{2}\widetilde{q}^{2}$ and
simple integration gives for $\omega <c\widetilde{q}_{0}$%
\begin{equation}
\chi _{2D}(\omega )=\frac{\overline{F_{\mathbf{q}}}}{4\pi c^{2}}\left[
\begin{array}{c}
\Phi (\omega ,\Delta _{\mathbf{M}},\gamma );\quad \quad for\emph{\;}\theta <1
\\
\pi +\Phi (\omega ,\Delta _{\mathbf{M}},\gamma );\quad for\emph{\;}\theta >1%
\end{array}%
\right]  \label{fi_from_hi}
\end{equation}%
\begin{eqnarray*}
\theta &=&(c^{2}\widetilde{q}_{0}^{2}+\Delta _{\mathbf{M}}^{2}-\omega
^{2})(\omega ^{2}-\Delta _{\mathbf{M}}^{2})/\gamma ^{2}\omega ^{2} \\
\Phi &=&\arctan \left\{ \frac{c^{2}\widetilde{q}_{0}^{2}\gamma \omega }{%
\gamma ^{2}\omega ^{2}+(c^{2}\widetilde{q}_{0}^{2}+\Delta _{\mathbf{M}%
}^{2}-\omega ^{2})(\Delta _{\mathbf{M}}^{2}-\omega ^{2})}\right\}
\end{eqnarray*}%
Here $\overline{F_{\mathbf{q}}}$ is the averaged smooth function $F_{\mathbf{%
q}}$.

In the limit under consideration the scaling denominator $\chi _{2D}(\omega
,T\rightarrow 0)$ is almost constant in a wide $\omega $-range and scaling
is ruled by $\chi _{2D}(\omega ,T)$. Accepting in (\ref{fi_from_hi}) linear $%
\gamma =\alpha T$ and $\Delta _{\mathbf{M}}=\beta T$ one obviously gets the
scaling ($\chi _{2D}(\omega ,T)$ becomes $\chi _{2D}(\omega /T)$). So in the
mentioned approximations the scaling function can be written as%
\begin{equation}
\widetilde{f}(\frac{\omega }{T})=\pi \Theta ((\frac{\omega }{T})^{2}-\beta
^{2})+\arctan \left\{ \frac{\alpha (\omega /T)}{(\beta ^{2}-(\omega /T)^{2})}%
\right\}  \label{scal_approx}
\end{equation}%
In contrast to numerous experimental fittings by simple $\arctan $, the
scaling function $\widetilde{f}(\omega/T)$ (\ref{scal_approx}) is described
by 'switched' $\arctan $ law and contains a microscopic information on $%
\Delta _{\mathbf{M}}$ and $\gamma $. The switching by $\Theta $-function
takes place at $\omega =\Delta _{\mathbf{M}}$.

In Fig.2 $\widetilde{f}$ is represented for $\alpha =0.25,\beta =0.5 $ by
dotted line $\mathbf{e}$ and it coincides with $f_{T=0.2,\gamma =0.05}$. The
scaling function $\widetilde{f}(\omega /T)$ with slightly different
parameters $\alpha =0.25,\beta =0.43$ is almost indistinguishable from
experimental $f_{ex}(\omega /T)$ \cite{Keimer92}. Let us remind, that in the
above calculations we have taken $\gamma \sim T$ and such $\lambda (T) $\
that the self-consistent calculations led to $\Delta \sim T$.

So in the dielectric limit (small $p$) the model leads to an adequate
description of experimental results. The scaling law strongly restricts $%
\gamma (T)$ dependence.

Now we turn to the case $p=0.28$ which corresponds to $\Delta _{\mathbf{M}%
}\approx \Delta _{\mathbf{X}}=\widetilde{\omega }_{\mathbf{q=X}}$. We relate
this case to the optimal doping. Calculated data are presented for $T=0.025$
and $0.05$\ with$\ $ $\gamma =0.38+0.8T$\ (in contrast to low frustration
limit here $\gamma $ does not tend to zero at $T\rightarrow 0$) and $\lambda
=10.0^{1/3}$. For $T=0.025$ and $0.05$\ \ the gaps are\ $\Delta _{\mathbf{M}%
}=0.197,$ $\Delta _{\mathbf{X}}=0.179$ and$\ \Delta _{\mathbf{M}}=0.228,$ $%
\Delta _{\mathbf{X}}=0.210.$

\begin{figure}
\includegraphics[width=8cm]{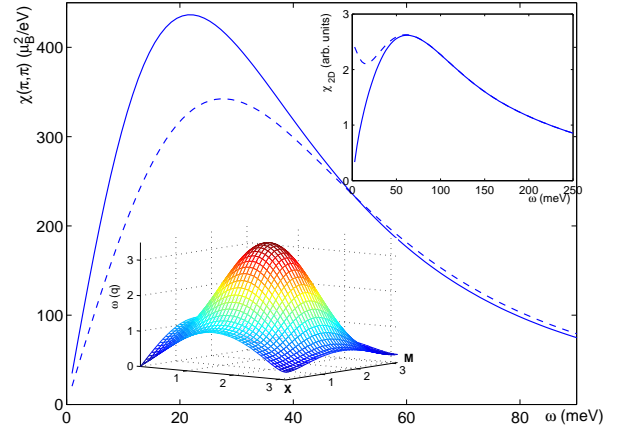}
\caption {$\chi ^{^{\prime \prime }}(\mathbf{Q},\omega )$ for
$T/I=0.025$ (solid) and $T/I=0.05$ (dashed curve) {\it Inset}:
$\chi _{2D}(\omega )$ -- solid line and $\chi _{2D}(\omega
)(2n_{Bose}+1)$ -- dashed line for $T/I=0.05$; {\it Lower inset}:
The spectrum $\widetilde{\omega }_{\mathbf{q}}/I$ for $p=0.28$,
$T/I=0.05.$} \label{fig:fig3}
\end{figure}

Fig.3 shows the $Q$-peaks, i.e. $\chi ^{^{\prime \prime }}(\mathbf{Q},\omega
),$\ for $T=0.025$\ and $T=0.05$. They are also in good agreement with the
experiment \cite{Fong2000}. In the inset of Fig.3 the calculated spectrum $%
\widetilde{\omega }_{\mathbf{q}}$ is shown for $T=0.05$. The dispersion $%
\widetilde{\omega }(\mathbf{q})$ has the following new features: the saddle
points close to $\mathbf{q=}(\pi ;\pi /2);(\pi /2;\pi )$ and $\widetilde{%
\omega }(\mathbf{q})$ changes weakly along $\mathbf{X-M}$\ direction. That
is why $\chi _{2D}(\omega )$\ has a peak at $\omega \gtrsim 2\div 3{\,}%
\Delta _{\mathbf{M}}$. This is explicitly seen in the another inset of Fig.3
which gives $\chi _{2D}(\omega )$\ (solid line)\ and $\chi _{2D}(\omega
)(2n_{Bose}+1)$\ (dashed line)\ for $T=0.05$. These curves qualitatively
correspond to the experimental ones \cite{cmr04} for optimally doped
curates. It is clear that the shown behavior of $\widetilde{\omega }(\mathbf{%
q})$ and $\chi _{2D}(\omega )$\ is a result of a stripe scenario if we
remind that the increase of $p$ drives the system to a state which is close
to a coherent superposition of two semiclassical stripe phases with $\Delta
_{\mathbf{X}}=0$\ \cite{Berezovsky}. It can be shown that $c_{\mathbf{q}%
}=\left\langle S_{\mathbf{q}}^{z}S_{-\mathbf{q}}^{z}\right\rangle \ $is
qualitatively different for small and large frustrations. For $p\leq 0.1$\
the structure factor $c_{\mathbf{q}}\ $has an extremely narrow peak at $%
\mathbf{q=Q}$.\ For $p=0.28$ the structure factor has peaks at $\mathbf{q=Q=M%
}$ and at $\mathbf{q=X}$. With $p$ increase the peaks at $\mathbf{X}$\
points increase and the $\mathbf{M}$ -peak disappears.

We capture this physics taking a spin-only model. But this model is too
simple to reflect a well-known low-energy incommensurate magnetic
excitations at wave vectors close to $\mathbf{Q}$\ at optimal doping. It is
obvious that one needs to introduce explicitly the spin-hole scattering to
describe this feature.

To check the applicability of the obtained spin susceptibility $\chi (%
\mathbf{q},\omega )$ for the kinetics of the optimally doped HTSC we
calculate the in-plane resistivity $\rho (T)$ and the Hall coefficient $%
R_{H}(T)$ in the framework of the spin-fermion model with the Hamiltonian
\begin{eqnarray}
\hat{H} &=&\hat{H}_{0}+\hat{H_I}  \label{Kin_ham} \\
\hat{H}_{0} &=&\sum_{\mathbf{k},\sigma }\varepsilon _{\mathbf{k}}a_{\mathbf{k%
}\sigma }^{\dagger }a_{\mathbf{k}\sigma }+J\frac{1}{\sqrt{N}}\sum\limits_{%
\mathbf{k},\mathbf{q},\gamma _{1},\gamma _{2}}a_{\mathbf{k}+\mathbf{q}%
,\gamma _{1}}^{\dagger }S_{\mathbf{q}}^{\alpha }\hat{\sigma}_{\gamma
_{1}\gamma _{2}}^{\alpha }a_{\mathbf{k},\gamma _{2}}  \notag
\end{eqnarray}%
The hole spectrum $\varepsilon _{\mathbf{k}}$ is obtained from the
calculation of the lower spin-polaron band in a six pole approximation \cite%
{Bar2001} and is shown in the inset of Fig.4.

It is well known that scattering by the spin-fluctuations with momenta $%
\mathbf{Q}$ leads to a strongly $T$-dependent anisotropy. To take it into
account the equation of motion for the non-equilibrium density matrix $\hat{%
\rho}^{(1)}=Z^{-1}\exp (-\hat{H}_{0}/T)\hat{F}$\ is solved by seven-moment
approach $\hat{F}=\sum_{l=1\div 7}\eta _{l}\hat{F}_{l}$, $\hat{F}_{l}=\sum_{%
\mathbf{k},\sigma }F_{l}(\mathbf{k)}a_{\mathbf{k}\sigma }^{\dagger }a_{%
\mathbf{k}\sigma }$. The moments $F_{l}(\mathbf{k})$ are taken to be
polynomials in velocity components $\mathbf{v_{k}}=\partial \varepsilon _{%
\mathbf{k}}/\hbar \partial \mathbf{k}$ and their derivatives: $F_{l}^{E}(%
\mathbf{k})=\{v_{\mathbf{k}}^{x},{\,}(v_{\mathbf{k}}^{y})^{2}v_{\mathbf{k}%
}^{x},{\,}\frac{\partial v_{\mathbf{k}}^{x}}{\partial y}v_{\mathbf{k}}^{y},{%
\,}\frac{\partial v_{\mathbf{k}}^{y}}{\partial y}v_{\mathbf{k}}^{x},{\,}%
\frac{\partial v_{\mathbf{k}}^{x}}{\partial x}\frac{\partial v_{\mathbf{k}%
}^{y}}{\partial y}v_{\mathbf{k}}^{x},{\,}(v_{\mathbf{k}}^{x})^{3}{,}\frac{%
\partial v_{\mathbf{k}}^{x}}{\partial x}v_{\mathbf{k}}^{x}\}.$ The detailed
expressions for $\rho (T)$ and $R_{H}(T)$ are given in \cite{Bar2006}. The
susceptibility $\chi (\mathbf{q},\omega )$ (\ref{hi_our}) is involved in
scattering integrals. To clarify the importance of the form (\ref{hi_our})
we also present the results for widely used so-called overdamped
susceptibility $\chi _{ovd}(\mathbf{q},\omega )$ (when $\omega ^{2}$ term in
the denominator of (\ref{hi_our}) is omitted) \cite{Pines, Hlubina}.

\begin{figure}
\includegraphics[width=8cm]{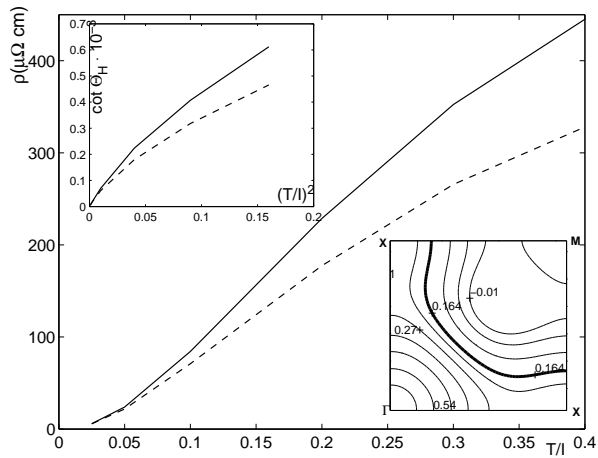}
\caption
{The resistivity $\rho (T/I)$ and cotangent of the Hall angle $cot{\,}\Theta
_{H}(T^{2}/I^{2})$ at 10 T ({\it upper inset}). The solid lines -- for $\chi (\mathbf{q},\omega )$ (\ref{hi_our}). The dashed lines -- for overdamped $\chi _{ovd}(\mathbf{q},\omega )$. {\it Lower inset}: the spectrum $\varepsilon _{\mathbf{k}}$ (in ${eV}$) given by the curves $\varepsilon _{\mathbf{k}}=\mathrm{const}$; bold curve -- Fermi line for optimal doping.}
\label{fig:fig4} \end{figure}

The results presented in Fig.4 are obtained for $p=0.28$, $I=100$ $meV$ and $%
J=200$ $meV$. The plots are the resistivity $\rho (T)$ and the Hall angle
cotangent $\ cot{\,}\Theta_{H}=\rho _{xx}/(R_{H}B)$ (in the inset) obtained
for the $\chi (\mathbf{q},\omega )$ (\ref{hi_our}) -- solid lines and for $%
\chi _{ovd}(\mathbf{q},\omega )$ -- dashed lines. In accordance with the
experiment \cite{Ando04}, the $\rho (T)$ curve exhibits a temperature
dependence close to a linear one starting from low $T$ with the value $\rho
(400K)/\rho (100K)$ $\approx 5$. It can be shown that $\chi _{ovd}(\mathbf{q}%
,\omega )$ approximation underestimates the scattering for large $\omega $.
As a result at $\rho (T)_{ovd}<\rho (T)$ and, as it is seen from Fig 4, in
some temperature regions $\rho (T)_{ovd}$\ has a different curvature. The$\
cot{\,}\Theta_{H}$ exhibits nearly linear behavior on $T^{2}$\ in a wide
temperature range, however, at low temperatures deviation from linearity is
obvious. It seems hopeful that the self-consistent spin susceptibility $\chi
(\mathbf{q},\omega )$ allows to describe experimental temperature anomalies
of two kinetic coefficients simultaneously.

In summary, we have made a systematic self-consistent study of the spin
problem in 2D frustrated Heisenberg antiferromagnet. Key features of the
model -- temperature dependence of the damping in low frustration limit and
the appearance of saddle points of the dispersion $\widetilde{\omega }(%
\mathbf{q})$ close to $\mathbf{q=}(\pi ;\pi /2);(\pi /2;\pi )$ in the case
of strong frustration increase -- allow to relate the results to a wide hole
doping interval in cuprates.

Work was supported by Russian Fund of Fundamental Investigations and Russian
Science Support Foundation.


\begin{thebibliography}{99}
\bibitem{Lee2006} P. A. Lee, N. Nagaosa, and X.-G. Wen, Rev. Mod. Phys.
\textbf{78}, 17 (2006).

\bibitem{Birgeneau06} R. J. Birgeneau \textit{et al.}, cond-mat/0604667
(2006).

\bibitem{Lake99} B. Lake \textit{et al.}, Nature \textbf{400}, 43 (1999).

\bibitem{Keimer92} B. Keimer \textit{et al.}, Phys. Rev. B \textbf{46},
14034 (1992).

\bibitem{Prelovsek04} P. Prelovsek, I. Sega and J. Bonca, Phys. Rev. Lett.
\textbf{92} 027002 (2004).

\bibitem{Larionov05} I. A. Larionov, Phys. Rev. B \textbf{72}, 094505 (2005).

\bibitem{Sherman03} A. Sherman, and M. Schreiber, Phys. Rev. B \textbf{68},
094519 (2003).

\bibitem{st91} H. Shimahara, and S. Takada, J. Phys. Soc. Jpn. \textbf{60},
2394 (1991).

\bibitem{Berezovsky} A. F. Barabanov, and V. M. Berezovsky, Phys. Lett. A
\textbf{186}, 175 (1994); JETP \textbf{79}, 627 (1994); J. Phys. Soc Jpn. 63
(1994) 3974.

\bibitem{Bar95} A. F. Barabanov, L. A. Maksimov, Phys. Lett. \textbf{A207},
390 (1995).

\bibitem{Hayden96-PRL} S. M. Hayden \textit{et al.},\textit{\ }Phys. Rev.
Lett. \textbf{76}, 1344 (1996).

\bibitem{Hayden96-PR} S.M. Hayden, G. Aeppli, T.G. Perring, H.A. Mook, F.
Dogan, Phys. Rev. B \textbf{54}, R6905 (1996).

\bibitem{Coldea01} R. Coldea \textit{et al.}, Phys. Rev. Lett. \textbf{86},
5377 (2001).

\bibitem{Stock04} C. Stock \textit{et al.}, Phys. Rev. B \textbf{69}, 014502
(2004).

\bibitem{cmr04} N. B. Christensen \textit{et al.}, Phys. Rev. Lett. \textbf{%
93}, 147002 (2004).

\bibitem{Fong2000} H. F. Fong \textit{et al.}, Phys. Rev. B \textbf{61},
14773 (2000).

\bibitem{Bar2001} A. F. Barabanov \textit{et al.}, JETP \textbf{92}, 677
(2001).

\bibitem{Bar2006} A. M. Belemouk, A. F. Barabanov and L. A. Maksimov ,
Zh.Eksp.Teor.Fiz. \textbf{129}, 493 (2006) [JETP \textbf{102}, 431 (2006)].

\bibitem{Pines} B. P. Stojkovic and D. Pines, Phys.Rev. B \textbf{55}, 8576
(1997).

\bibitem{Hlubina} R. Hlubina, and T. M. Rice, Phys. Rev. B \textbf{51}, 9253
(1995).

\bibitem{Ando04} Y. Ando, S. Komiya, K. Segawa, S. Ono, Y. Kurita , Phys.
Rev. Lett. \textbf{93}, 267001 (2004).
\end{thebibliography}
\end{document}